# On A New Formulation of Micro-phenomena: Quantum Paradoxes And the Measurement Problem

Afshin Shafiee[1]
Research Group On Foundations of Quantum Theory and Information,
Department of Chemistry, Sharif University of Technology,
P.O.Box 11365-9516, Tehran, Iran.

**Abstract**

As a serious attempt for constructing a new foundation for describing micro-incidents from a local causal standpoint, I explained before that each micro-entity can be assumed to be composed of a probability field joined to a particle (called a particle-field or PF system) which is endowed with energy, having an objective character. In this essay, some odd quantum phenomena including the tunneling effect, the measurement problem and the EPR theorem are reexamined in detail from the standpoint of the new theory.

## 1 Introduction

Quantum mechanics is a mathematical tool which enables us to make some predictions about the probability of obtaining different outcomes in repeated experiments. This theory provides with us nothing about the individual results, even at the observational level of perception. It challenges our intuitive grasp of micro-events and still in a broader world-view, it stimulates counterintuitive attitudes regarding the meaning of the physical reality, the significance of causality and the role of observer in the measuring process. Nevertheless, there is no inclusive consensus among the physicists and the philosophers of science about the meaning of quantum theory and the way one is preferred to look at quantum world. So, to clarify the perplexing sides of the quantum description about the micro-incidents, there have been suggested many interpretations since the formulation of quantum theory to increase the explanatory power of it. The interpretation of the quantum world, however, is not unique and unrivaled. All interpretations are to some extent adequate and tolerable in some definite domains of description and at the same time deficient and fragile in some other contexts. The most fragile point in any interpretation of quantum mechanics is that the interpreter tries to *explain* any micro-physical incident according to the *descriptive* power of the quantum theory. This is indeed a failure to keep eyes partially open towards what we can learn from Nature.

Yet, the door for exploring new ideas is still open [1]. Nevertheless, there are many weird quantum phenomena which all call for a *coherent* physical explanation in any alternative approach. The most important of them are the measurement problem, the EPR paradox [2] and the quantum interference phenomenon described by the famous double-slit experiments [3, 4]. The EPR paradox has kept involved people's attention towards either the so-called nonlocal realistic alternatives (such as the Bohm theory [5]), or local antirealistic interpretations of quantum mechanics (recently manifested more radically in some information-based interpretations [6, 7]) for more than 70 years. Regarding the measurement problem, on the other side, the decoherency approach is among the most powerful formalisms given so far for solving this problem [8, 9]. The decoherence formalism, however, gives us no clear depiction of the micro-entities (e.g., the meaning of dual wave-particle character of quantum systems), it does not explain the nature of the EPR-type correlations and adds nothing to our knowledge more than what the standard quantum mechanics describes about the reality of physical quantities before measurement. The epistemic disposition of such approaches goes no farther from the scopes of standard quantum theory itself, although each renovated approach surely sheds new light on the meaning of quantum theory to a certain extent.

---

[1]E-mail: shafiee@sharif.edu

As a result, no generally accepted variant of quantum mechanics has been provided up to now to explain all the above puzzling phenomena in a coherent physical basis. In our terminology, a "coherent physical basis" signifies a merged physical ground consisting of general rules and regularities with an explanatory capability not restricted to a specific epistemic domain. Correspondingly, a "coherent physical theory" should explain why different regularities appear in distinctive physical contexts, how various theories could be reconciled on a common consistent ground and even how one can explore new realms of physics unknown thus far based on the same common basis. While all these requirements may primarily seem inaccessible and unattainable, the search for finding new alternative formulations should be seriously continued.

To be involved in such an important effort, we have attempted to construct a new foundation for describing micro-events from a deterministic causal standpoint. It has been explained before in the first essay of this series that in our new approach, one encounters an unified concept of information, matter and energy [1]. Here, each micro-particle is assumed to be enfolded by a probability field which is endowed with energy, having an objective feature. Each particle-field (called a PF system) satisfies a Newton-like equation of motion influenced by a complicated force which its form can be only recognized only after knowing what fields are the related solutions of the Schrödinger equation. The physical status of each individual PF system is also influenced by the whole character of the total system including other particles in a local context-dependent way. So, beyond the mechanical interactions a given particle is assumed to be directly subjected to, the very existence of a *collection* of interacting particles might cause the state of each PF system be affected by the presence of other particles, even though there would be no direct interaction between all of them.

In the sequel of the previous work, here, we intend to survey some weird quantum phenomena from the standpoint of our new theory. This article is composed of the following parts. In section 2, tunnelling effect will be studied in our approach to explain how a particle can pass through a potential barrier, when its total energy is smaller than the energy of barrier. In section 3, we talk about the measurement problem from a realistic particle-field standpoint. This section includes physical details regarding the momentum and position measurements in micro-domain. Then, in section 4, the EPR problem is analyzed to show how what EPR introduces as a completeness criterion is fulfilled in our theory. At the end, in section 5, a summary of the whole content of this paper is provided.

## 2   Tunneling Effect

For a one-dimensional system in the $x$-direction, we consider a particle with energy $E < V_0$ incident on a potential barrier of height $V_0$ and width $a$, that is $V = V_0$ for $0 \leq x \leq a$. The potential energy of particles is zero before and after the barrier. Using the time-independent

Schrödinger equation, for $x < 0$, the stationary field of the system is given by

$$\chi_I = Ae^{ikx} + Be^{-ikx} = R_I \cos(kx + \vartheta_I) \qquad (1)$$

where $A = \frac{R_I}{2}\exp(i\vartheta_I)$, $B = \frac{R_I}{2}\exp(-i\vartheta_I)$, $k = \sqrt{\frac{2mE}{\hbar^2}}$ and $R_I$ and $\vartheta_I$ are two real constants. The subscript $I$ denotes the region $x < 0$. Here, $\chi_I$ is interpreted as a transverse field and $R_I$ has a dimension of length. At $x < 0$, the particle is free of any potential, so we have:

$$\frac{d^2\chi_I}{dt^2} = -\omega_I^2 \chi_I \qquad (2)$$

where $\omega_I = v_P^{(I)} k$ and $v_P^{(I)}$ is the velocity of particle in the first region $I$. Using the above equation, one can simply obtain the kinetic and potential energies of the field as

$$K_{F,I} = \frac{1}{2}m\omega_I^2 R_I^2 \sin^2(kx + \vartheta_I) \qquad (3)$$

and

$$V_{F,I} = \frac{1}{2}m\omega_I^2 R_I^2 \cos^2(kx + \vartheta_I) \qquad (4)$$

respectively. The energy of field is then equal to:

$$E_{F,I} = \frac{1}{2}m\omega_I^2 R_I^2 = K_{P,I} k^2 R_I^2 \qquad (5)$$

where $K_{P,I}$ is the kinetic energy of particle in the first region. Thus, one can express the total energy of the system as:

$$E = K_{P,I}(1 + k^2 R_I^2) \qquad (6)$$

Similarly, for the particle in the region $0 \leq x \leq a$, its associated stationary field can be represented by:

$$\chi_{II} = Ce^{\kappa x} + De^{-\kappa x} = R_{II}\cosh(\kappa x + \vartheta_{II}) \qquad (7)$$

where $\cosh(u) = \frac{\exp(u) + \exp(-u)}{2}$. Also, $C = \frac{R_{II}}{2}\exp(\vartheta_{II})$ and $D = \frac{R_{II}}{2}\exp(-\vartheta_{II})$ are real constants which again $R_{II}$ has a dimension of length. The subscript $II$ denotes the region $0 \leq x \leq a$ and $\kappa = \sqrt{\frac{2m(V_0 - E)}{\hbar^2}}$. The equation of motion of the field (7) is given by

$$\frac{d^2\chi_{II}}{dt^2} = \omega_{II}^2 \chi_{II} \qquad (8)$$

where $\omega_{II} = v_P^{(II)}\kappa$ and $v_P^{(II)}$ is the velocity of particle inside the barrier. Interestingly, in the second region the sign of the spring-like force becomes positive, indicating that at $0 \leq x \leq a$, there should exist an attracting potential energy corresponding to a total negative energy of field. In other words, in this region the particle takes energy from its surrounding field to increase its kinetic energy.

Strictly speaking, the potential energy of the field inside the barrier can be given by

$$V_{F,II} = -\frac{1}{2}m\omega_{II}^2 R_{II}^2 \cosh^2(\kappa x + \vartheta_{II}) \tag{9}$$

where the minus sign indicates that opposite to the barrier potential $V_0$, there is an attracting potential $V_{F,II}$ which lowers the whole potential of the PF system. Using the relation $K_F = \frac{1}{2}m|\dot{\chi}|^2 = \frac{1}{2}mv_P^2|\chi'|^2$ (where " dot" and " prime" denotes $\frac{d}{dt}$ and $\frac{d}{dx}$, respectively), the kinetic energy of field can be expressed by

$$K_{F,II} = \frac{1}{2}m\omega_{II}^2 R_{II}^2 \sinh^2(\kappa x + \vartheta_{II}) \tag{10}$$

From the relations (9) and (10), the energy of field is given by:

$$E_{F,II} = -K_{P,II}\kappa^2 R_{II}^2 \tag{11}$$

Then, one can obtain the total energy of the PF system as:

$$E = K_{P,II}(1 - \kappa^2 R_{II}^2) + V_0 \tag{12}$$

which has the same value as (6). The relation (12) shows that the whole PF system actually experiences not a barrier potential $V_0$ alone, but a reduced value of potential energy $(V_0 - K_{P,II}\kappa^2 R_{II}^2)$.[2] This means that when the particle goes into the classically forbidden region, it exchanges energy with its associated field in a manner that the kinetic energy of the particle $K_{P,II}$ increases, decreasing the energy of its surrounding field $E_{F,II}$, so that the total value of the potential term $(V_0 - K_{P,II}\kappa^2 R_{II}^2)$ decreases, but the total energy remains conserved. In effect, the particle takes energy from its surrounding to pass the barrier.

Considering the relations (6) and (12) and with a little manipulation, one can show that

$$E = \frac{K_{P,I}}{1 - \frac{p_{P,I}^2 R_I^2}{\hbar^2}} \tag{13-a}$$

and

$$E = \frac{K_{P,II}}{1 - \frac{p_{P,II}^2 R_{II}^2}{\hbar^2}} + V_0 \tag{13-b}$$

---

[2] We recall from [1] again that a PF system is in fact a particle, but with renovated physical properties brought about by the presence of a field associated and joined with the particle. This is the main difference between a PF system and a classical particle for which all physical properties are delineated for the very particle itself. In the classical domain, there is defined no other -associated- physical entity with which the particle shares its energy.

Since $E > 0$, relation (13-a) shows that

$$p_{P,I}^2 R_I^2 < \hbar^2 \tag{14}$$

where $p_{P,I}$ is the linear momentum of particle before the barrier and

$$R_I = \pm \frac{\hbar}{p_{P,I}} \left(1 - \frac{K_{P,I}}{E}\right)^{1/2} \tag{15}$$

Relation (14) says that for particles with high velocity, the magnitude of the field's amplitude is reduced. As the kinetic energy of the particle grows, the particle's energy $K_{P,I}$ approaches the total energy $E$, so that the energy of field will diminish. At the limit $E \to K_{P,I}$, the field will have no contribution in the total energy of the PF system, because $R_I \to 0$.

On the other hand, remembering that $E < V_0$, one can deduce from (13-b) that

$$p_{P,II}^2 R_{II}^2 > \hbar^2 \tag{16}$$

where $p_{P,II}$ is the linear momentum of particle inside the barrier. Here, if one of the values of $p_{P,II}$ or $R_{II}$ decreases, the other value should increase considerably.

One can find that the amplitude of field inside the barrier is given by

$$R_{II} = \pm \frac{\hbar}{p_{P,II}} \left(1 + \frac{K_{P,II}}{V_0 - E}\right)^{1/2} \tag{17}$$

Assuming that $E < V_0$, the amplitude of the field inside the barrier never comes near to zero and the high values of the kinetic energy of particle is not allowed too. For the latter case, if we assume that $p_{P,II}^2$ increases considerably, one can show that $R_{II} \to \pm \frac{\hbar}{\sqrt{2m(V_0-E)}} = \pm \frac{1}{\kappa}$, so that in (12) $E \to V_0$. But, this is in contrast to our earlier assumption that $E < V_0$. Accordingly, the kinetic energy of the particle inside the barrier should have a finite value, so that the energy contribution of the field cannot be neglected at all. In other words, the tunneling effect has no classical explanation, presuming merely a particle-based dynamics.

The expectation value of the kinetic energy in quantum formalism is negative inside the barrier. This can be also illustrated in relation (13-b) by defining

$$K_{II} = \frac{K_{P,II}}{1 - \frac{p_{P,II}^2 R_{II}^2}{\hbar^2}} = E - V_0 \tag{18}$$

where $K_{II}$ is the so-called kinetic energy in quantum mechanics (albeit, when both $E$ and $V_0$ are constant). We have discussed it before that when the energy of field is negative and its absolute value is greater than the kinetic energy of particle, the quantum mechanical definition of the linear momentum becomes meaningless [1]. However, this does not mean that there is no kinetic energy definable for the system. This only means that the description of the standard quantum mechanics is inadequate in such baffling situations.

# 3 The Measurement Problem

In this section we are going to scrutinize one of the most important issues in the micro-domain, namely the measurement problem. To elucidate the problem, we carefully look over two special cases, i.e., linear momentum and position measurements which -we think- are sufficiently subtle issues to contain the main key points.

In many text books, a measurement process is introduced as a state preparation stage followed by a detection step [10]. For example, suppose that an electron is sent from a source and scattered before reaching a detection screen. The scattering process can be described as a preparation stage followed by detecting the position of electron on a definite location of screen. In a different procedure, however, for measuring the position of the scattered electron, one may also consider another preparation stage with an alteration in the operation of the screen. That is, after the scattering of electron and before reaching the screen, we can first prepare a *position eigenstate* of the system by setting a series of narrow slits along the screen. After each slit a detector is placed. Of course, it cannot be predicted with certainty in what slit the electron is going to land. But, a little after the electron arrived in one of the slits, it would be detected by an detector and its position will be revealed. In both of these events, we finally measure the position of the electron, but in the second one, we prepare the state of the system in a *definite* position eigenstate at a given time before detecting it. While in the first setup, these two events take place at the same time, i.e., we detect the position of the electron and by the very action of detection we prepare a position eigenstate too. Physically and practically, both of the above procedures describe the same phenomenon, but in the latter case, the relationship between the measurement and the preparation of a relevant eigenstate becomes more evident. For this reason, it is sometimes said that a system of slits is a position-measuring device [11].

From a fundamental point of view, we define the "measurement" as a physical process in which a physical property (in micro-domain, an observable) becomes a *discernible* (i.e., both distinguishable and perceivable) quantity in practice. In classical physics, measurement is indeed a passive action, for any physical property is a discernible quantity by itself. Notwithstanding, in micro-world not every real property is always perceivable. Accordingly, the measurement process has an essential role for a conscious observer to gain knowledge about definite discernible values of a property in practice for a given system.

In our approach to the description of micro-phenomena, a discernible quantity can be defined based on the equivalence of what is observed for an observable in practice and its corresponding value for a given PF system. As we stated earlier [1]: "[w]henever the mean value of a quantity in a given eigenstate is the same value obtained for the PF system, one can assert that it is a discernible property in practice." On the other hand, such an equivalence takes place only during a measurement process in which (as explained above) a definite eigenstate of an observable is prepared.

## 3.1 The linear momentum measurement

To measure the linear momentum of a PF system, one should first put the system into a physical situation in which the *particle* is subjected to no mechanical force whatever. Accordingly, we prepare a momentum field which is free of any interaction too. To prepare a stationary field $\chi$, in general, it is sufficient the particle is being subjected to a conservative force $f_P$ [1]. Then, considering a one-dimensional system for simplicity, the force $f_F$ the field experiences can be

given by

$$f_F = mv_P^2 \frac{d|\chi'|}{dx} + f_P|\chi'| \tag{19}$$

where $v_P$ is particle's velocity and $\chi' = \frac{d\chi}{dx}$. The stationary $\chi$ functions satisfy the time-independent Schrödinger equation. So, if $f_P = 0$ everywhere in space (resulting from the fact that the potential energy of particle $V_P = 0$), $\frac{d|\chi'|}{dx}$ will be zero too and hence $f_F = 0$.

Let us assume that at $t < t_0$, the particle has been subjected to a non-conservative force. We denote its associated non-stationary field as $X(x(t), t)$. The non-stationary field satisfies both the time-dependent Schrödinger equation (derived for any field in [1]) as well as the Newton-like equation

$$m \frac{d|\dot{X}|}{dt} = f_{nc,F} \tag{20}$$

where $f_{nc,F}$ is a non-conservative force and

$$\dot{X} = \left(\frac{\partial X}{\partial x}\right) v_P(t) + \left(\frac{\partial X}{\partial t}\right); \quad t < t_0 \tag{21}$$

At $t = t_0$, we release the particle from its preceding force and let it be free of any force whatever. That is

$$m \frac{d|\dot{\chi}|}{dt} = 0 \tag{22}$$

In effect, noticing that $f_P = f_F = 0$ in (22) (see also (19)) which means that we have a free PF system at $t = t_0$ and later on, for $t \geq t_0$ we have:

$$\chi_p(x(t)) = A_p \exp(ikx(t)); \quad t \geq t_0 \tag{23}$$

where $A_p$ is the amplitude of the momentum field and $k = \frac{p}{\hbar}$ which $p$ is the de Broglie momentum. Also, $x(t)$ is the position of particle at $t \geq t_0$. The energy of field at $t \geq t_0$ is equal to the kinetic energy $K_F$, because the potential energy of field $V_F$ is zero. Thus, using (23), we get:

$$E_F = K_F = \frac{1}{2} m v_P^2 k^2 A_p^2 = \frac{p_P^2}{2m} k^2 A_p^2 \tag{24}$$

where $p_P$ is the linear momentum of the particle which is constant at $t \geq t_0$. Noticing that $V_P = 0$ and remembering that in general $E = V_P + \left(E_F + \frac{p_P^2}{2m}\right) = V_P + \frac{p^2}{2m}$ [1], one can show that

$$\frac{p}{m} = v_P (1 + k^2 A_p^2)^{1/2} \tag{25}$$

which has the same value of the velocity of the PF system $\dot{q}$ at $t \geq t_0$, where

$$\dot{q} = \dot{x}\left(1 + \frac{|\dot{x}_p|^2}{v_P^2}\right)^{1/2} = v_P\left(1 + k^2 A_p^2\right)^{1/2} \tag{26}$$

Since the momentum of the PF system is equal to the mean value of the linear momentum $\langle p \rangle_{\chi_p} = \hbar k$ at $t \geq t_0$, one can assign a *discernible* momentum to the system which can be detected in practice. So, we have described the physical mechanism by which the de Broglie momentum becomes a discernible quantity during a measurement process.

Regarding the relation (25), one can show that

$$p = p_P\left(1 - \frac{p_P^2 A_p^2}{\hbar^2}\right)^{-1/2} \tag{27}$$

from which one can conclude that

$$p_P^2 A_p^2 < \hbar^2 \tag{28}$$

In the inequality (28), it is obvious that as the momentum of particle becomes larger, the amplitude of the momentum field becomes smaller, so that for the classical objects, it is anticipated that $A_p \to 0$. Considering the definition of the de Broglie momentum $p = \frac{h}{\lambda}$ and using the relation (27), we can obtain a relation for the associated wavelength of the particle as

$$\lambda = \frac{h}{p_P}\left(1 - \frac{p_P^2 A_p^2}{\hbar^2}\right)^{1/2} \tag{29}$$

which shows that as $p_P$ increases and $A_p$ decrease, $p_P^2 A_p^2$ remains a definite value smaller than $\hbar$, but $\lambda$ approaches zero because of the factor $\frac{h}{p_P}$.

The trajectory of the PF system after the preparation of a momentum eigenstate can be given by integrating the relation (26) over time:

$$q(t) = \frac{p}{m}(t - t_0) + q_0; \quad t \geq t_0 \tag{30}$$

where $p$ is given by (27) and

$$q_0 = x_0\left(1 - \frac{p_P^2 A_p^2}{\hbar^2}\right)^{-1/2} \tag{31}$$

which $x_0$ is the position of particle at $t = t_0$.

It should be noted that the underlying physics of the linear momentum measurement is a *reversible* process, although at two different stages $t < t_0$ and $t \geq t_0$, the field experiences two different forces $f_{nc,F}$ and $f_F = 0$ in (20) and (22), respectively, corresponding to two different forces $f_{nc,P}$ and $f_P = 0$ the particle is subjected to. The dynamics of the PF system changes at $t_0$ due to the change in the forces the PF system experiences, but the latter dynamics begins continuously after the former one. So, we have a sudden but continuous change in the trajectory of system at $t = t_0$. Strictly speaking, using the equation (30), one can infer the value of $q_0$, having the value of the $q(t)$ at $t$. By knowing the value of $q_0$ at $t_0$, also, it was possible *in principle* to infer the value of $q(t)$ at $t < t_0$, if we could know the force $f_{nc,F}$ in (20) determined by having the preceding non-stationary field $X(x(t),t)$.

At last, it should be pointed out that according to what our formalism describes, no reduction takes place in transition from a physical situation characterized by the physical state of a PF system to another one described by another state. Taking into account the physical foundation of particles and their associated field, the quantum wave functions can be introduced satisfying the time-dependent and time-independent Schrödinger equations [1]. We have argued before that the square of the absolute value of each wave function represents a probability density, so $|X|^2$ is a probability field. There is always a dual correspondence between the fields and wave functions under the linear regime of field equations, because the function-dependency of them is the same. So, in a similar way, to each quantum eigenfunction, there corresponds an eigenfield. Since each set of quantum eigenfunctions forms a complete set, the same relation holds true for their corresponding set of eigenfields. Accordingly, it is possible to expand a desirable field in terms of a set of eigenfields at a proper time. In this way, one can relate two different fields $X$ and $\chi_p$ defined above at $t_0$, according to the following equation:

$$X(x_0, t_0) = \int_{-\infty}^{+\infty} dp \ \chi_p(x(t_0))\phi(p, t_0) \tag{32}$$

where $x_0 = x(t_0)$. In (32), we have considered all possible fields $\chi_p$, corresponding to all possible values of $p$ which one can attribute to a given PF system at $t = t_0$.

Using the relation (32), one can obtain the mean value of the momentum of a PF system, considering all possible values it can possess in an experiment:

$$\langle p \rangle_{t_0} = \frac{1}{(2\pi\hbar)A_p^2} \int_{-\infty}^{+\infty}\int_{-\infty}^{+\infty}\int_{-\infty}^{+\infty} dx_0 dp dp' \ \chi_p^*(-i\hbar\frac{\partial}{\partial x_0})\chi_{p'}\phi^*(p, t_0)\phi(p, t_0)$$
$$= \int_{-\infty}^{+\infty} dp \ p|\phi(p, t_0)|^2 \tag{33}$$

where $p$ is introduced in (27) and $\langle p \rangle_{t_0}$ is the mean value of the momentum of a PF system, albeit considering all possible values of $p$.[3] The above relation shows that $|\phi(p, t_0)|^2$ should be interpreted as a probability density. It says that *if* at $t = t_0$ one measures the momentum of the PF system, the probability of observing a value of $p$ between $p$ and $p + dp$ will be given by $|\phi(p, t_0)|^2$. The probability density $|\phi(p, t_0)|^2$ is a *contingent* probability, contingent upon measuring the momentum of a PF system at $t = t_0$. If one really measures the momentum of a PF system at $t = t_0$, the associated field of the particle will be $\chi_p(x(t_0))$ at that time, not $X(x_0, t_0)$, and the PF system will possess a definite value of $p$. However, for knowing the exact value of $p$, one should be aware of the particle's momentum $p_P$ as well as the field's amplitude $A_p$ in (27) which are both hidden parameters. So, different values of $p$ might be observed in experiment. To each definite determinate value of $p$, however, there corresponds a specific $\chi_p$. Thus, for an ensemble including particles all with the same associated fields $\chi_p$, the mean value $\langle p \rangle_{\chi_p}$ is equal to an exact value $p$ attributed to each PF system in that ensemble. Nevertheless, if we consider a more extensive set of results comprised of all possible values of $p$ that a given PF system can possess in similar experiments (or, alternatively, an ensemble including the PF systems with different possible values of $p$), the mean value of the momentum will be given by (33). No reduction happens in (32) at $t = t_0$.

---
[3]See also Appendix A in [1].

## 3.2 The position measurement

To measure the position of a PF system, we should first prepare conditions in which PF's position approaches the position of an unfolded particle, not allied with an energetic field, at a given time, i.e., $q \to x$ (we work only again in one dimension for simplicity). Suppose that at $t < t_0$, there is a particle with an arbitrary allied field $X(x(t), t)$ corresponding to the wave function $\Psi(x(t), t)$. To measure its position, we should *find* the PF system at $t = t_0$ within a very narrow slit along the $x$-direction, where the slit is located at $a \leq x \leq a + \delta a$ and $\delta a \approx 0$. After the PF system passed through the slit, its position (here, e.g., $x \approx a$) could be discerned by detecting the *particle* immediately after the slit.

What does happen, when the PF system enters the slit at $t = t_0$? The slit looks like a one-dimensional infinite potential energy well with width $\delta a$. Rearranging the spatial coordinate of slit from $x$ to $x - a$ in the $x$-direction, one can suppose that the potential energy is zero at $0 \leq x - a \leq \delta a$ and infinite at boundaries. Thus, when the PF system arrives at the slit, the particle experiences new conditions. Within the slit, the PF system is released from its surrounding interactions. Hence, $f_{P,x} = 0$ at $t = t_0$, where $f_{P,x}$ is the force exerted on particle along the $x$-direction. In such conditions, the equations of motion of particle and its associated field are determined by the following relations, respectively

$$m \frac{d^2 x(t)}{dt^2} = 0; \quad t = t_0 \tag{34}$$

and

$$\frac{d^2 \chi(x(t))}{dt^2} = -\bar{\omega}^2 \chi(x(t)); \quad t = t_0 \tag{35}$$

where $\bar{\omega}^2 = v_P^2 k^2$, $v_P$ is the velocity of particle and $k^2 = \frac{2mE}{\hbar}$. Since $f_{P,x} = 0$ within the slit, the energy of the PF system $E$ is conserved at $t = t_0$. From the equation (35), one can find that

$$\chi(t) = A_0 \sin(\bar{\omega} t + k x_0); \quad t = t_0 \tag{36}$$

where $\bar{\omega} = k v_p$ and $A_0$ is the amplitude of $\chi$ and $x_0 = x(t_0)$ is particle's position at $t_0$. Taking advantage of the methods developed in [1], one can also find the position of the PF system at $t = t_0$ as:

$$q(x - a) = (x - a) + \delta a F_k(x); \quad t = t_0 \tag{37}$$

For a quantized stationary field $\chi = \chi_n(x)$, the function $F_{k_n}(x)$ can be read out as [1]:

$$F_{k_n}(x) = \frac{1}{n\pi}\{b_{n1}\sin[2k_n(x-a)] - b_{n2}\sin[4k_n(x-a)]+\ldots\} \qquad (38)$$

where $k_n = \frac{n\pi}{\delta a}$ and $b_{n1}$, $b_{n2}$, ... are some coefficients which their definitions is not important here.[4]

It is now apparent from (37) that because $\delta a \approx 0$, so $q(x) \approx x$ and $\dot{q}(x) \approx v_P$. Thus, the amplitude $A_0$ in (36) should come near to zero (hence, the energy of field, i.e., $E_F \approx 0$) and $E \approx E_P = \frac{1}{2}mv_P^2$. In this way, the particle becomes unfolded at $t = t_0$ with no energetic field allied with it. Concurrently, the particle's position $x \approx a$ can be detected at $t = t_0$.

At $t < t_0$, the trajectory of the PF system is given by $q(x(t), t)$ which includes some hidden parameters, such as the velocity or position of the particle and the amplitude of its associated field. On the other hand, as stated before for the linear momentum measurement, the dynamics of the PF system changes suddenly at $t_0$ due to the changes in the forces the particle and field experience in consequence of the new conditions imposed on the system. But, this is not a continuous change for the position measurement. Some amount of energy should be released or taken by the particle at $t = 0$, since the energy of its allied field (which might be negative or positive before measurement) is vanished at that time. This energy transfer between the particle and its associated field is not dynamical, because there is no mechanical interaction between them, but is indeed an irreversible conversion process, leading to an unfolding particle at the time of measurement.

Accordingly, the dynamics of the PF system suddenly *breaks* at $t = t_0$ and we have a discontinuity in the trajectory of system, because at $t = t_0$ we have not a PF system, but a particle only with its own properties. By knowing the position of the *particle* at $t_0$ (i.e., $x_0$), one cannot *in principle* determine the position of the *PF system* $q_0$ at that time (but only a formal function $q(x_0, t_0)$ can be inferred), even if all other hidden parameters could be known simultaneously. Strictly speaking, in general, $q_0 \neq x_0$. Hence, in contrast to the linear momentum case, the position measurement is an *irreversible* process. By knowing $x_0$ at $t = t_0$, we cannot determine the position of the PF system at any prior time. There is no continuous trajectory imaginable for a micro-*particle*.

In effect, one encounters two different problems here. Firstly, due to the hidden character of $q(x(t), t)$, we cannot determine with certainty at what specific location the *PF system* could be *present* at $t = t_0$, if even no measurement took place at that time. Secondly, one cannot also predict with certainty where the *particle* can be *found*, when its position is measured at $t_0$, even if he or she could know the value of $q(x(t), t)|_{t=t_0} = q_0$. Measurement process in indeed a physical intervention which changes the circumstances each micro-entitiy (comprised of a particle and its associated field) is subjected to. In the case of the position measurement, one sees even a more concrete intervention, because it causes a shift in the position of a micro-entity during the measurement.

Because of the probabilistic nature of wave functions, however, one can evaluate the probability that the particle can be observed within a given range of space, when its position is measured. To do this, one should first note that the probability of observing the particle between $a \leq x \leq a + \delta a$ in a one-dimensional space is nearly one for $\delta a \approx 0$. So, while the amplitude of particle's field is nearly zero in a narrow slit, the spatial distribution of the particle can be given by a sharp Dirac delta function $\delta(x_0 - a)$ and the position of particle becomes a discernible property.[5] Correspondingly, $\langle x_0 \rangle_{Df} = a$ ($Df$ stands for Dirac function) for all cases in which the same value of $a$ has been observed for the particle in different runs of experiment (or, alternatively, in an ensemble of particles all with the same location $a$ at $t = t_0$). Yet, as explained above, the value of $a$ is unpredictable. So, one may consider a more extensive set of results including all possible values of $a$ detected in similar experiments (or, alternatively, an ensemble including the particles with different possible values of $a$). Then, the mean value of the particle's position at $t_0$ is given by

$$\langle x_0 \rangle_{t_0} = \int_{-\infty}^{+\infty} dx_0 \ x_0 |\Psi(x_0, t_0)|^2 \tag{39}$$

where $\Psi(x_0, t_0)$ is the wave function of the PF system at $t = t_0$.[6]

---

[4] See Appendix B in [1] for more details.

## 4 The EPR thought experiment

In 1935, Einstein, Podolsky and Rosen (EPR) argued that presuming the reality of two incompatible physical quantities described by non-commuting operators (position and linear momentum, for instance), the description of reality in quantum mechanics is not *complete* [2]. In their argument, the requirement for a theory to be complete is introduced by a necessary condition stating that "every element of the physical reality must have a counterpart in the physical theory" which is called the *condition of completeness*. EPR, also, introduce a sufficient condition for the reality of physical properties which is defined as the possibility of predicting a physical quantity with certainty, without disturbing the system.

Besides, their reasoning was implicitly based upon two other assumptions, i.e., the locality and separability conditions. The first notion indicates that in a many-particle system, the properties of each particle do not depend on the measurements performed simultaneously on the other spatially separated particles. The second assumption (i.e., the separability condition) implies that one can establish the properties of the whole system based upon the properties of individual particles. Those who have faith in the completeness of quantum mechanics (albeit, in the framework of EPR's definition of reality), usually accentuate the nonlocal (or nonseparable) character of quantum phenomena, without clarifying what "quantum nonlocality" (or "quantum nonseparability") really means.

Adopting EPR's epistemic definition of the physical reality as a sufficient condition together with embracing the locality and separability criteria, we are going to show that our new description of micro-phenomena fulfills the requirement of completeness defined above.

Similar to EPR's original argument and using the usual terminology of quantum mechanics, let us consider a pair of particles described by the following wave function in configuration space (here in one-dimension for the present):

$$\psi(x_1, x_2) = \int_{-\infty}^{+\infty} dp\, u_p(x_1) w_p(x_2) \tag{40}$$

where $p$ denotes the linear momentum and $x_1$ ($x_2$) is the position variable of particle 1 (2). Also, one can define the following momentum eigenfunctions for particles 1 and 2, respectively:

$$u_p(x_1) = \frac{1}{\sqrt{2\pi\hbar}} \exp\left(\frac{ix_1 p}{\hbar}\right); \quad w_p(x_2) = \frac{1}{\sqrt{2\pi\hbar}} \exp\left[\frac{-i(x_2 - x_0)p}{\hbar}\right] \tag{41}$$

where $x_0$ is a constant with dimension of length. According to (40) and using the momentum eigenfunctions defined in (41), if we measure the momentum of the first particle and obtain a definite value $p_1 = p$ at a given time $t = 0$, we can predict with certainty that the momentum of the second particle should be $p_2 = -p$ at the same time. Thus, in accordance with the criterion of reality, we must consider the quantity $p_2$ as being an element of reality.

---

[5] Interestingly, in the classical limit in which the probability distributions of conservative systems are changed to their corresponding classical ones, we can show that the field's amplitude becomes zero, so the particle facet will dominate. For, in the limit that the classical distributions describe the statistical behavior of the system, the difference between the total energy and particle's energy disappears and the energy of field becomes zero. Thus, we have probability distributions for which no energetic field can be delineated.

[6] We recall that from the energy pattern of the whole PF system before measurement, we can qualitatively predict at what locations the detection of particle is more (less) probable. Maximum and minimums of the kinetic energy of the PF system is in accordance with the minimums and maximums of the square of stationary wave functions (or particle's stationary fields), respectively, showing that in general, the square of the absolute value of any wave function $|\Psi|^2$ can be interpreted as a probability density.

On the other hand, using the above definitions, the relation (40) can be rewritten as

$$\psi(x_1, x_2) = \delta(x_1 - x_2 + x_0) \tag{42}$$

where $\delta$ denotes Dirac delta function. The above relation can be expanded in terms of the position eigenfunctions of particles 1 and 2, denoted by $v_x(x_1)$ and $\varphi_x(x_2)$, respectively:

$$\psi(x_1, x_2) = \int_{-\infty}^{+\infty} dx\, v_x(x_1) \varphi_x(x_2) \tag{43}$$

where

$$v_x(x_1) = \delta(x - x_1); \quad \varphi_x(x_2) = \delta(x - x_2 + x_0) \tag{44}$$

Thus, if we were going to measure the position of the first particle at $t = 0$, we would obtain a given value $x$. Correspondingly, we could attribute the value of $x + x_0$ to the position of the second particle at the same time. Hence, we should consider the quantity $x_2$ as being an element of reality as well at $t = 0$, receiving the fact that the reality of $p_2$ and $x_2$ does not depend on the process of measurement carried out on the first particle. Because, the two particles are *separated* enough that no measurement on one of them can influence the other simultaneously. Therefore, EPR conclude that two incompatible quantities (here, the position and momentum of the second particle) are simultaneously real and the quantum description of physical reality given by the wave function $\psi(x_1, x_2)$ is not complete.

Now, we analyze the EPR thought experiment in our approach. Let us assume that at $t = 0$, the state $\psi(x_1(0), x_2(0))$ is prepared:

$$\psi(x_1(0), x_2(0)) = \delta(x_1(0) - x_2(0) + x_0) \tag{45}$$

where $x_1(0)$ ($x_2(0)$) is the position of the first (second) particle at $t = 0$ and $x_1(0) - x_2(0) = x(0) = -x_0$. The whole system includes two PF systems (i.e., two particles with their associated fields) which are propagated in space without any interaction with each other or with their environment, so that the linear momentum of the whole system is conserved at any instant. Subsequently, at any later time $t > 0$, the entire state of system can be represented by a time-dependent field $X(x(t), t)$ where $x(t) = x_1(t) - x_2(t)$. Describing the time evolution of the field in terms of an appropriate free-particle propagator $K(x(t), t; x(0), 0)$ (which is an alternative but equivalent approach for the time-dependent Shrödinger equation), one can write:

$$X(x(t), t) \propto \int_{-\infty}^{+\infty} dx(0) K(x(t), t; x(0), 0) \delta(x(0) + x_0)$$
$$= A(t) t^{-1/2} \exp\left[\frac{im(x(t)+x_0)^2}{2\hbar t}\right] \tag{46}$$

where $A(t)$ is a coefficient depending on time, $m$ is the mass of each particle and

$$K(x(t), t, -x_0, 0) = \sqrt{\frac{m}{i\hbar t}} \exp\left[\frac{im(x(t)+x_0)^2}{2\hbar t}\right] \tag{47}$$

Using the Newton equation for the pair particles, it is simple to show that

$$x(t) = \frac{p_P}{m} t + x(0) \tag{48}$$

where $p_P = p_{P_1} - p_{P_2}$ and $p_{P_1}$ ($p_{P_2}$) is the linear momentum of the first (second) particle. Taking into account the relation (48), the field relation (46) can be represented as $X(x(t), t) \equiv X_{p_P}(x(t), t)$, where

$$X_{p_P}(x(t), t) = A(t) t^{-1/2} \exp\left(\frac{i p_P x(t)}{2\hbar}\right) \exp\left(-\frac{i p_P x(0)}{2\hbar}\right)$$
$$= c_F \exp\left(\frac{i p_P x(t)}{2\hbar}\right) \exp\left(-\frac{i p_P x(0)}{2\hbar}\right) \tag{49}$$

Here, based on the no-physical-interaction hypothesis, we have used the equality $A(t) = c_F \sqrt{t}$ where $c_F$ is a constant, having the dimension of length (see Appendix A). Thus, in (49), one can define $X_{p_P}(x(t), t) = \chi_{p_P}(x(t))$.

In relation (46), It is important to note that at $t \to 0$, we have two equivalent representations of $X_{p_P}(x(0), 0)$, but with two different mathematical representations. On one hand, taking into account the free propagator relation in (47), one can show that:

$$X_{p_P}(x(0), 0) = A(0) \sqrt{\frac{i\hbar}{m}} \delta(x_1(0) - x_2(0) + x_0) \tag{50}$$

which is a null field, because $A(0) = 0$. On the other hand, considering the relation (49), we have

$$X_{p_P}(x(0), 0) = c_F \exp\left(\frac{i p_P x_1(0)}{2\hbar}\right) \exp\left(-\frac{i p_P (x_2(0) - x_0)}{2\hbar}\right) \tag{51}$$

which is equivalent to (50), only if we take $c_F = 0$. This means that, according to the relation (45), when we prepare a two-*particle* state at $t = 0$, no energetic field will be associated with each particle at any later time $t > 0$ as well. Nevertheless, the field representations at $t = 0$ are crucial, since different function-dependencies in (50) and (51) demonstrate the real essence of both particles' position and momentum properties as is required by a complete physical theory to describe them.

The field relation (49) can be factorized based on the individual fields $\chi_{p_P}^{(1)}(x_1(t))$ (for particle 1) and $\chi_{p_P}^{(2)}(x_2(t))$ (for particle 2), that is

$$\chi_{p_P}(x(t)) \propto \chi_{p_P}^{(1)}(x_1(t))\chi_{p_P}^{(2)}(x_2(t)) \tag{52}$$

where

$$\chi_{p_P}^{(1)}(x_1(t)) = c_{F_1}\exp\left(\frac{ipx_1(t)}{\hbar}\right) \tag{53-a}$$

and

$$\chi_{p_P}^{(2)}(x_2(t)) = c_{F_2}\exp\left(-\frac{ip(x_2(t)-x_0)}{\hbar}\right) \tag{53-b}$$

Here, $c_{F_1}$ and $c_{F_2}$ are the amplitudes of the first and second particles' allied fields, respectively, both equal to zero here, and $p = \frac{p_P}{2}$. Also, considering the relation (27), one can similarly show that

$$p = p_{P_1}\left(1 - \frac{p_{P_1}^2 c_{F_1}^2}{\hbar^2}\right)^{-1/2} = p_{P_1} \tag{54}$$

and

$$-p = p_{P_2}\left(1 - \frac{p_{P_2}^2 c_{F_2}^2}{\hbar^2}\right)^{-1/2} = p_{P_2} \tag{55}$$

From the relations (51), (54) and (55), it is obvious that one can assign two discernible momentums $p_{P_1} = \frac{p_P}{2}$ and $p_{P_2} = -\frac{p_P}{2}$ to the first and second particles, respectively, similar to what described before in (23). In other words, the individual fields (53-a) and (53-b) are plane waves describing definite values of momentum.

One can plainly demonstrate that the trajectories of the pair particles at $t \geq 0$ satisfy the following relations (see relation (30) for comparison):

$$x_1(t) = \frac{p}{m}t + x_1(0); \quad x_2(t) = -\frac{p}{m}t + x_2(0) \tag{56}$$

where $p_{P_1} = p = \langle p \rangle_{\chi_{p_P}^{(1)}}$ and $p_{P_2} = -p = \langle p \rangle_{\chi_{p_P}^{(2)}}$.

Based on the relations (52) and (53-a), if we measure the momentum of particle 1 at any given time $t \geq 0$, we will get a result $p = p_{P_1}$. Then, the second particle's momentum will be definitely $p_{P_2} = -p$, as can be inferred from (53-b). The latter quantity thus fulfills the epistemic definition of the physical reality given by EPR. The momentum of the first particle $p_{P_1}$ is also real (because the particle possesses it independent of any observation), but cannot be predicted with certainty, because the trajectory of particle 1 (described by $x_1(t)$) cannot be discerned in practice. Yet, at $t = 0$, we have shown that the joint field of the two particles can be represented as two equivalent relations (50) and (51). From (50), one can predict the second particle's position with certainty, knowing the position of the first particle at $t = 0$, since $x_2(0) = x_1(0) + x_0$. On the other hand, from (51), one can predict the second particle's momentum, knowing the momentum of the first particle at the same time, because $p_{P_2} = -p_{P_1} = -\frac{p_P}{2}$. Thus, in accordance with EPR's criterion of reality, we can consider both incompatible quantities $x_2(0)$ and $p_{P_2}$ as being the elements of reality. Since these two elements of physical reality have explicit counterparts in our physical approach, the underlying theory is *complete* as well.

Similar to what explained before in section 3.1, it is possible to relate the functions of relations (50) and (51) to each other (see the relation (32), for instance). If the two particles are separated at $t = 0$ with a definite distance $x_2(0) - x_1(0) = x_0$, the initial wave function of system will be given by (45), in conformity with its field representation (50). However, while the value of $p_P$ in (51) is predeterminate for each specific joint particles, there is no way we know it beforehand, for the values of $p_{P_1}$ and $p_{P_2}$ cannot be assigned in advance. Accordingly, considering an ensemble including a plenty of two-particle systems comprised of all possible values of $p = \frac{p_P}{2}$, one can write:

$$\psi(x_1(0), x_2(0)) = \frac{1}{2\pi\hbar} \int_{-\infty}^{+\infty} dp \ \exp\left(\frac{ipx_1(0)}{\hbar}\right) \exp\left(-\frac{ip(x_2(0)-x_0)}{\hbar}\right) \tag{57}$$

where $\psi(x_1(0), x_2(0))$ has been defined in (45) and we have used the fact that the momentum eigenfunctions form a complete set. The relation (57) is similar to the relation (40) in quantum description, but with a time constraint at $t = 0$.

The relations (52) and (56) shows that the underlying description of the two-particle system is consistent with the assumptions of locality and separability. For, the whole physical state of the system can be constructed upon the properties of individual particles which are not influenced by any action made on their mates.

Moreover, if one analyzes the EPR experiment in a three-dimensional space where any local observer can measure the position or the momentum of a given particle along arbitrary directions at the same time, one can obviously show that the underlying field representation of the two-particle system in (52) can be also generalized to contain all directions. It is because the components of position (or momentum) are compatible observables, so we need not different field relations corresponding to different measuring contexts. Accordingly, the relation (57) can be written as:

$$\psi(\vec{r}_1(0), \vec{r}_2(0)) = \frac{1}{(2\pi\hbar)^3} \int_{-\infty}^{+\infty} d^3p \ \exp\left(\frac{i\vec{p}\cdot\vec{r}_1(0)}{\hbar}\right) \exp\left(-\frac{\vec{p}\cdot(\vec{r}_2(0)-\vec{r}_0)}{\hbar}\right) \tag{58}$$

where $\vec{r}_j$ is the local vector of the $j$th particle ($j = 1, 2$) in the Euclidean space and $\psi(\vec{r}_1(0), \vec{r}_2(0)) = \delta(\vec{r}_1(0) - \vec{r}_2(0) + \vec{r}_0)$. Also, $\vec{p} = -\vec{p}_{P_2} = \vec{p}_{P_1}$ is the momentum vector determinable in practice. From (58), the relation (57) can be obtained as a special case. Hence, there is no room for context-dependency in the original EPR thought experiment.

Interestingly, it was also noted by Bell that the existence of a non-negative Wigner distribution for the original EPR state enables us to provide a local classical model for describing the position and momentum correlations [12]. His statement has been critically assessed by many authors in recent years (see, e.g., [13]), by proposing counterexamples in which an EPR state is used to yield a violation of Bell's inequality [14]. However, the subtle point is that in all reported works in this regard, what is really proposed is an experimental scheme for observing the correlations between some *discrete* observables (other than the continuous position and momentum variables) including some incompatible ingredients which allow for the violation of a Bell-type inequality. Thus, although the EPR state (42) is *utilized* in calculations, the violation of the Bell-type inequality is demonstrated in a discrete Hilbert space for observables which their corresponding group is *not* Abelian.

Similarly, the so-called no-go theorems (such as the Bell theorem [15]) usually contain discrete variables like the components of the spin variable which are not compatible observables in different directions. Thus, apparently, the situation depicted above for the original EPR entangled state cannot be approved in physical situations in which there are limitations in measuring different components of an observable at the same time. In such circumstances, it is not possible to write down a general relation like (58) from which one can infer the relationships between all components of a given property for a joint two-particle system concurrently. In this regard, we have recently demonstrated that the *spin* correlation of two entangled particles in a singlet state can be reproduced by a local *contextual* hidden-variable model which shows that the notion of contextuality may enter in other circumstances besides the original EPR case [16]. What can be definitely said here is that the original EPR proposal can be explained in our approach, approving the locality and separability as well as the context-independency criteria. Yet, this is not a general rule for all entangled states, when different measuring contexts are not compatible for some given pair of observables, corresponding to a non-Abelian group (like the SU(2) group for the Pauli spin operators).

## 6   Conclusion

In this paper, we discussed about three important topics in quantum theory. We first examined the tunneling effect and showed how the negative energy of the field inside the barrier lessens the potential energy of the barrier and enables the particle passes through it. Here, the existence of an undivided notion of particle-field helps us again to resolve one the best-known puzzles in quantum domain on a rigorous physical basis.

The measurement problem was the second topic we inspected here. It is really one of the most significant problems in quantum mechanics which has been always a subject of debate in literature. We explained the measurement problem on the basis of the appearance of a *discernible* quantity. A discernible quantity makes sense, when suitable physical conditions are made ready for its appearance. To see the linear momentum of a micro-entity, we should release the *particle* from its surrounding interactions. This causes that a stationary field corresponding to a definite value of momentum is prepared, although until we have not observed what momentum the PF system possesses, we are ignorant of its exact value. What really happens here is a continuous change of dynamics of the PF system, according to what we prepare for the system. Thus, no reduction happens for a given field and the fields before and after the change of dynamics can be always related to each other at the time of alteration.

The position measurement has generally the same description as above. However, there is a peculiar feature in position measurement which makes it to be distinguished from any other measuring process. In any spatial measuring setup, we always measure the position of a *particle*, not the PF system. This fact causes the dynamics of the system breaks at the moment of the measurement, because two different physical objects are under consideration before and during the measurement. In a position measurement, we again set free the PF system from any interaction at the time of measurement, however this time not in the entire space, but within a small location. This makes the PF system be unfolded as a particle in a definite local space. So, the PF's dynamics changes to a new status in which the preceding properties should be redefined for a particle only, at the moment (say $t = 0$) a position eigenstate is prepared. This is a discontinuous change, because if, e.g., the position of the PF system is predicted to be $q_0$ at $t = 0$, the real position of system is $x_0$ belonging to a definite location of *particle*. The change from $q_0$ to $x_0$ is not a dynamic change, but is an alteration in conditions the system experiences at that time. Corresponding to the negative or positiveness of the energy of particle's associated field, some amount of energy is released or taken by the particle at $t = 0$, respectively. This energy transfer between the particle and its allied field causes the particle be unfolded at the time of measurement. So, the position of a PF system is always hidden. We can just measure the particle's location discontinuously.

Regarding the original EPR theorem, we showed how in our approach both the position and momentum of a particle can be simultaneously described at a hidden level. The fields describing these two conjugate variables for a joined two-particle system possess zero amplitudes. Hence, we have null fields encompassing two mathematical representations in (50) and (51). This enables us to describe the incompatible observables of the EPR experiment in a consistent manner. Whenever we have null fields with zero amplitudes but well-defined functional representations, we have merely lack of knowledge about the real properties of system and our ignorance about the factual states of affairs is a subjective matter only. Thus, our theory fulfills the completeness criterion in EPR theorem. Nevertheless, as far as the entanglement notion is introduced in spatial coordinates (like the original EPR thought experiment), the underlying description of problem in our approach is consistent with the assumptions of locality and separability and is context-independent. Context-dependency, on the other hand, is an essential feature of our formalism which disappears in special circumstances, when, e.g., there is no limitation in measuring different components of an observable at the same time [1]. This is indeed the situation we encounter in the EPR proposal. The situation differs, however, when spin correlations are proposed. The meaning and formulation of spin is still an open problem in our approach, but an instructive consistent model has been recently provided which can reproduce the spin correlation of a singlet Bell state in a local-contextual manner [16]. In the framework of our formalism, this may be considered as a basis for formulating a more coherent model of spin in future.

**Acknowledgment**

It is a pleasure to thank Abouzar Massoudi, Farhad T. Ghahramani, Iman Khatam and Zahra Mashhadi for their stimulating discussions about the subject of the paper. My thanks also go to my students Mohammad Bahrami, Majid Karimi, Arash Tirandaz, Hossein Sadeghi, Keivan Sadri and Shobeir K. Seddigh for their comments and suggestions during the preparation of the first version of the manuscript.

**Appendix A**

To find the field coefficient $A(t)$ in relation (49), we use the physical point that the two-particle system and its associated field are free of any interaction during the separation of the particles from the source. Thus,

$$\frac{d|\dot{X}_{p_P}|}{dt} = 0 \tag{A-1}$$

where, using (49), we can show that

$$\dot{X}_{p_P} = \left[-\frac{A(t)}{2t^{3/2}} + \frac{\dot{A}(t)}{t^{1/2}} + \frac{ip_P^2}{2m\hbar}\frac{A(t)}{t^{1/2}}\right] \exp\left(\frac{ip_P x(t)}{2\hbar}\right) \exp\left(-\frac{ip_P x(0)}{2\hbar}\right) \tag{A-2}$$

So, we get:

$$|\dot{X}_{p_P}| = \frac{1}{t^{1/2}}\left[\left(\dot{A} - \frac{A}{2t}\right)^2 + A^2\left(\frac{p_P^2}{2m\hbar}\right)^2\right]^{1/2} \tag{A-3}$$

Now, putting (A-3) in (A-1), we obtain:

$$\left(\dot{A} - \frac{A}{2t}\right)^2 + A^2\left(\frac{p_P^2}{2m\hbar}\right)^2 = 2t\left(\dot{A} - \frac{A}{2t}\right)\left(\ddot{A} - \frac{\dot{A}}{2t} + \frac{A}{2t^2}\right) + 2A\dot{A}t\left(\frac{p_P^2}{2m\hbar}\right)^2 \tag{A-4}$$

Equalizing the terms with same factors in (A-4), one gets:

$$\left(\dot{A} - \frac{A}{2t}\right) = 2t\left(\ddot{A} - \frac{\dot{A}}{2t} + \frac{A}{2t^2}\right); \quad A = 2\dot{A}t \tag{A-5}$$

from which it is straightforward to show that $A = c_F\sqrt{t}$, where $c_F$ is a constant.

## References

[1] A. Shafiee, "On A New Formulation of Micro-phenomena: Basic Principles, Stationary Fields And Beyond", quant-ph/arXiv:0810.1031.

[2] A. Einstein, B. Podolsky, and N. Rosen, "Can Quantum-Mechanical Description of Physical Reality Be Considered Complete?" *Physical Review*, **41**, 777-780 (1935).

[3] R. P. Feynman R. B. Leighton and M. Sands, *The Feynman Lectures on Physics*, (Addison-Wesely, Reading, 1965), Vol. 3.

[4] A. Shafiee, A. Massoudi and M. Bahrami, "On A New Formulation of Micro-phenomena: The Double-slit Experiment", quant-ph/arXiv:0810.1034.

[5] D. Bohm and B. J. Hiley, *The Undivided Universe*, (Routledge, London, 1993).

[6] A. Zeilinger, "A Foundational Principle for Quantum Mechanics", *Foundations of Physics*, **29**, 631-643 (1999).

[7] A. Shafiee, F. Safinejad and F. Naqsh, "Information And The Brukner-Zeilinger Interpretation of Quantum Mechanics: A Critical Investigation", *Foundations of Physics Letters*, **19**, 1-20 (2006).

[8] W. H. Zurek, "Environment-induced Superselection Rules", *Physical Review D*, **26**, 1862-1880 (1982).

[9] W. H. Zurek, "Decoherence, Einselection And The Quantum Origin of The Classical", *Reviews of Modern Physics*, **75**, 715-775 (2003).

[10] L. E. Ballentine, *Quantum Mechanics* (World Scientific, New Jersey, 2001), p. 230.

[11] C. Cohen-Tannoudji, B. Diu and F. Laloe, *Quantum Mechanics* (Wiley, New York, 1977), p. 261.

[12] J. S. Bell, "EPR correlations and EPW distributions", *Annals of New York Academy of Sciences*, **480**, 263 (1986).

[13] A. F. Abouraddy, T. Yarnall, B. E. A. Saleh and M. C. Teich, "Violation of Bell's inequality with continuous spatial variables", *Physical Review A*, **75**, 052114 (2007).

[14] J. S. Bell, "On the Einstein-Podolsky-Rosen paradox", *Physics*, **1**, 195-200 (1964).

[15] J. S. Bell, *Speakable and Unspeakable in Quantum Mechanics*, (Cambridge University Press, Cambridge, 1987).

[16] A. Shafiee, R. Maleeh and M. Golshani, "Common Cause and Contextual Realization Of Bell Correlation", *Annals of Physics*, **323**, 432--443 (2008).